\documentclass[%
 reprint,
 pre,
 amsmath,amssymb,
 aps,
longbibliography
]{revtex4-1}
\usepackage{amsmath}
\usepackage{mathbbol}
\usepackage{amssymb}             
\usepackage{amsthm}
\usepackage{siunitx}
\usepackage{graphicx}
\usepackage{bm}

\begin{document}

\newcommand{\beq}{\begin{equation}}
\newcommand{\eeq}{\end{equation}}
\newcommand{\derpar}[2]{\frac{\partial #1}{\partial #2}}
\newcommand{\eps}{\epsilon}
\newcommand{\dgam}{\dot\gamma}
\newcommand{\VEC}[1]{\mathbf{#1}}
\newcommand{\HAT}[1]{\hat{\mathbf{#1}}}
\newcommand{\dif}{{\rm d}}

\title{Non-ideal rheology of semidilute bacterial suspensions}

\author{Marcelo Guzm\'an}
\author{Rodrigo Soto}
\affiliation{Departamento de F\'\i sica, Facultad de Ciencias F\'\i sicas y Matem\'aticas, Universidad de Chile, 
Avenida Blanco Encalada 2008, Santiago, Chile}
\date{\today}

\begin{abstract}
The rheology of semidilute bacterial suspensions is studied with the tools of kinetic theory, considering binary  interactions, going beyond the ideal gas approximation. Two models for the interactions are considered, which encompass both the steric and short range interactions. In these, swimmers can either align polarly regardless of the state previous to the collision or they can align axially, ending up antiparallel if the relative angle between directors is large. In both cases, it is found that an ordered phase develops when increasing the density, where the shear stress oscillates with large amplitudes, when a constant shear rate is imposed. This oscillation disappears for large shear rates in a continuous or discontinuous transition, depending on if the aligning is polar or axial, respectively. For pusher swimmers these non-linear effects can produce an increase on the shear stress, contrary to the prediction of a viscosity reduction made for the dilute regime with the ideal gas approximation. 
\end{abstract}

\pacs{Valid PACS appear here}
\maketitle

\section{Introduction}
Bacterial suspensions have attracted great attention in recent years as one of the quintessential examples of active matter~\cite{berg2008coli,vicsek2012collective,marchetti2013hydrodynamics}. They exhibit striking features which are absent in their passive counterparts, such as collective oscillatory behavior~\cite{chen2017weak}, active turbulence~\cite{wensink2012meso,cisneros2007fluid}, hydrodynamic instabilities~\cite{saintillan2008instabilities,pahlavan2011instability}, and exotic rheological properties~\cite{gachelin2013non,lopez2015turning,sokolov2009reduction,rafai2010effective,saintillan2010dilute}  (for a recent review on the rheology, see Ref.~\cite{saintillan2018rheology}).
Mechanically, swimmers can be classified according to the sign of the force dipole they exert on the fluid as pushers for extensile dipoles and pullers for contractile ones.
Experiments on \textit{Escherichia coli} suspensions (pushers) revealed a decrease in the total viscosity  with the swimmer concentration~\cite{gachelin2013non,sokolov2009reduction}, reaching even ``superfluid'' states at moderate concentrations~\cite{lopez2015turning},  whereas experiments on \textit{Chlamydomonas reinhardtii} (pullers) exhibited a net increase on the viscosity~\cite{rafai2010effective}. These modifications on the suspension viscosity have been accounted for through kinetic theories~\cite{saintillan2010dilute} in the dilute regime, i.e. when all the swimmer--swimmer interactions are neglected in the ideal gas rheology. 
It is found that for small shear rates, the active contribution to the viscosity is proportional to the dipole strength and bacterial concentration, being negative for pushers and positive for pullers in accordance with the experiments. This effect  depends on the swimmer shape, vanishing for spherical swimmers and being maximal for elongated bodies. Finally, at larger shear rates, the active suspension becomes non-Newtonian~\cite{saintillan2010dilute,gachelin2013non}. 

The interactions among swimmers are far from simple, but they can be  classified as long- and short-range. The former, of hydrodynamical origin, leads, for instance, to instabilities for pushers~\cite{saintillan2008instabilities} and to spiral vortices in confinement~\cite{wioland2013confinement}. The latter, due to steric and hydrodynamic mechanisms,  leads to the formation of polar and nematic phases~\cite{cisneros2007fluid,sokolov2007concentration}, and flocking transitions~\cite{vicsek1995novel}. Density plays a major role in selecting which interaction dominates. At low concentrations, short-range interactions are negligible, while for moderate and high concentrations they take over the long-range hydrodynamics~\cite{cisneros2007fluid,drescher2011fluid}. This has been observed as an effective ``screening'' for very dense suspensions~\cite{wensink2012meso,marchetti2013hydrodynamics}. Finally, as the long-range hydrodynamic interactions are mediated through the divergence of the stress tensor~\cite{saintillan2008instabilities}, this regime appears whenever the characteristic length of the fluctuations is smaller than the average distance among bacteria. 
In particular, they can be discarded for homogeneous systems. 
Finally, long-range hydrodynamic interaction can be included with a mean field description, where the local rheology is governed by the short-range interactions~\cite{saintillan2008instabilities,guazzelli2011physical}.
In this article, we aim to characterize the non-ideal rheology of bacterial suspensions in the semi-dilute regime, hence, considering the effects of short-range interactions.

\section{Kinetic theory description}
To simplify the analysis and render more transparent the rheological response, we consider a bacterial suspension in two spatial dimensions, with conclusions that can directly be extended to three dimensions. Bacteria are described by their position $\VEC{r}=x\HAT{x}+y\HAT{y}$ and director, which is characterized by a single angle, $\HAT{p}=\cos\theta\HAT{x}+\sin\theta\HAT{y}$. To analyze the bacterial suspension, we use kinetic theory, which was previously used to quantify the viscosity reduction of a dilute suspension of microswimmers~\cite{saintillan2010dilute}. In this framework, the object under study is the distribution function $\Psi(\VEC{r},\theta,t)$, such that $\Psi\dif x\dif y\dif\theta$ gives the number of bacteria with the specified position and director. 
The suspension is placed in a uniform shear flow, with velocity $\VEC{u}=u(y)\HAT{x}=\dot\gamma y \HAT{x}$, where $\dot\gamma$ is the shear rate.

There are three contributions responsible for the change of the distribution function. First, bacteria move as an effect of the imposed flow and their self-propulsion with velocity $V_0$, which we assume to be constant and the same for all swimmers,  
\begin{align}
\dot{x} &=u(y)+V_0 \cos\theta, &
\dot{y} &=V_0 \sin\theta,
\end{align}
and the orientation changes as an effect of the local shear rate, described by the Bretherton--Jeffery equation
\begin{align}
\dot{\theta} &=\frac{\dot\gamma}{2}[\beta \cos (2\theta)-1],
\end{align}
where $\beta$ is the Bretherton coefficient, which for \textit{E. coli} can be taken as $0.7$, representing its elongated body. These deterministic equations of motion are transcribed to the kinetic equation as advection terms. The second contribution is the rotational diffusion with coefficient $D_\text{r}$, which is accounted for with a Fokker--Planck term. Tumbling, which also changes the orientations randomly, gives rise to rotational diffusion at long times, and its effect is absorbed into $D_\text{r}$. The explicit inclusion of tumbling with a Lorentz-like term in the kinetic equation~\cite{saintillan2010dilute,soto2016kinetic} alters only quantitatively the conclusions of this article. Translational diffusion is not considered, because the Brownian diffusivity is much smaller than the Berg diffusivity $D_\text{Berg}=V_0^2/D_\text{r}$~\cite{howse2007self}, and its inclusion results only in small corrections. 
Finally, in the semidilute regime, for homogeneous states where long-range hydrodynamic interactions can be neglected, bacteria have binary interactions.
These are accounted for with a Boltzmann-like bilinear collision term with both swimmers located at the same position.
Considering all these elements, the kinetic equation reads
\begin{equation}
\derpar{\Psi}{t}+\derpar{(\dot{x} \Psi)}{x} + \derpar{(\dot{y} \Psi)}{y} +  \derpar{(\dot\theta \Psi)}{\theta}=D_\text{r}\derpar{^2 \Psi}{\theta^2}+J,
\label{FPB}
\end{equation}
where $J$ is the collisional integral. Short-range steric \cite{peruani2006nonequilibrium,baskaran2008hydrodynamics,wensink2012meso} and hydrodynamic interactions \cite{ishikawa2006hydrodynamic,llopis2010hydrodynamic,lushi2014fluid} align bacteria. For simplicity,  we consider that the alignment is total, meaning that after their interaction, both swimmers emerge with the same angle, equal to the average of the incoming ones, resulting in 
\begin{equation}
J=g\int_{-\pi}^{\pi}dw\left[\Psi (\theta+w/2)\Psi(\theta-w/2)-\Psi(\theta)\Psi(\theta-w)\right],
\label{collisionalIntegral}
\end{equation}
as was introduced in Refs.~\cite{aranson2005pattern,aranson2007model} to describe the alignment of microtubules and  bacteria.
Here, $g$ quantifies the collision rate. In two dimensions, it scales as $g=\sigma_{2D}V_0$, where $\sigma_{2D}$ is the two-dimensional cross section. For \textit{E. coli}, we can use $V_0\approx\SI{20}{\micro\meter/\second}$ \cite{berg2008coli} and $\sigma_{2D}\approx\SI{2}{\micro\meter}$ to estimate $g\approx\SI{40}{\micro\meter^2/\second}$. Also, for \textit{E. coli},  $D_\text{r}\approx\SI{0.0125}{\second^{-1}}$~\cite{commentDr}.

The total shear stress is the sum of the viscous stress $\Sigma_{xy}^\text{v}=\eta\dot\gamma$, where $\eta$ is the fluid viscosity, and the particle stress. The latter is the  sum of the drag stress for passive suspensions $\Sigma_{xy}^\text{d}$ and the active contribution $\Sigma_{xy}^{\text{a}}$~\cite{saintillan2010dilute}. This last contribution, responsible for the effective viscosity reduction, is
\begin{align}
\Sigma_{xy}^{\text{a}}
=\sigma_0\int \Psi \cos\theta \sin\theta \dif\theta,
\end{align}
which can be computed once the solution of the kinetic equation is found. Here, $\sigma_0$ is the strength of the elementary force dipole, which is negative (positive) for pushers (pullers). For active suspensions, $\Sigma_{xy}^{\text{a}}$ dominates over $\Sigma_{xy}^\text{d}$, allowing us to discard the latter.

Equation \eqref{FPB} with $g=0$ (no binary collisions) has been solved in Ref.~\cite{saintillan2010dilute} finding the viscosity reduction described above. The opposite case, with interactions and $\zeta=0$,  presents a polar phase above a threshold density~\cite{aranson2005pattern}. In both limits, the  homogenous case already presents  interesting features that could be observed experimentally. In this article, we analyze the homogeneous suspension when both $g$ and $\zeta$ are finite, showing that intriguing rheological features appear.

\section{Rheological response}
Decomposing the distribution function in an angular Fourier series, $\Psi(\theta,t)=\sum_n a_n(t) e^{in\theta}$, with $a_{-n}=a_n^*$, the kinetic equation for a homogeneous suspension reads
\begin{align}
\frac{d b_n}{ds}=&-b_n\left(n^2-in\zeta/2+\rho \right)-i\zeta\beta n  \left( b_{n-2}+b_{n+2}\right)/4 \nonumber \\
&+2\pi \sum _{l} b_l b_{n-l}\text{sinc} \left(n\pi/2-l\pi\right),
\label{FPBn}
\end{align}
where $\text{sinc}(x)=\sin(x)/x$ and we employed the following rescaling to dimensionless variables: 
$s=D_\text{r}t$, $\zeta=\dgam/D_\text{r}$, and $b_n= a_n g/D_\text{r}$. 
The global dimensionless density is $\rho=2\pi b_0$, which is conserved, implying that $db_0/ds=0$.
In terms of the Fourier coefficients, the average orientation vector and the swimming contribution to the shear stress are
\begin{align}
\left<\HAT{p}\right>&=\frac{2\pi}{\rho} 
\begin{pmatrix}
         \text{Re}(b_1) \\
	-\text{Im}(b_1)
\end{pmatrix}, 
&
\Sigma^{\text{a}}_{xy}&=-\frac{\sigma_0\pi D_\text{r}}{g}\text{Im}(b_2).
\label{aSwimmingSS}
\end{align}

The rotational diffusion produces a fast decay of the high $n$ modes. It is possible then to truncate Eq.~\eqref{FPBn} with few modes, as the rest will take small values. Considering that the swimming contribution to the stress tensor depends on $a_2$, we first truncate the series to $n=2$ for the purpose of analytic calculations, 
\begin{align}
\frac{d b_1}{ds}&=b_1\left(\frac{i\zeta}{2}+\epsilon \right)-\frac{i\beta\zeta}{4}b_{-1}-\frac{8}{3}b_{-1}b_2, \label{eqa1}\\
\frac{d b_2}{ds}&=-\frac{i\beta\zeta\rho_c}{4\pi}\left( \epsilon+1\right)-\left[4-i\zeta+\rho_c(\epsilon+1) \right]b_2+2\pi b_1^2, \label{eqa2}
\end{align}
where we have defined the bifurcation parameter $\epsilon=\rho (4/\pi-1)-1$. 
When $\zeta=0$, an isotropic to polar transition takes place at $\epsilon=0$, that is, for the critical density $\rho_c =\pi/(4-\pi)\approx3.66$~\cite{aranson2005pattern}. 
Close to the transition, the growth rate of $b_1$ is much smaller than the decay rate of $b_2$, and therefore $b_2$ is enslaved to $b_1$. This analysis leads to the scalings $b_1\propto \sqrt{\epsilon}$ and $b_2\propto \epsilon$. 

The imposed shear flow when $\zeta\neq0$ induces a macroscopic rotation of $\HAT{p}$ and phase oscillations will appear in $b_1$ and $b_2$. For $|\epsilon|\ll1$ a multiple time scale approach is possible, where we define the fast and slow time scales $s_0=s$ and $s_1=\epsilon s$.
Based on the couplings in Eqs.~\eqref{eqa1} and \eqref{eqa2}, we propose the following ansatz:
\begin{align*}
&b_1(s_0,s_1)=Ae^{i\omega_0 s_0+i\omega_1s_1}+Be^{-i\omega_0 s_0-i\omega_1s_1},\\
&b_2(s_0,s_1)=Fe^{2i\omega_0 s_0+2i\omega_1s_1}+Ge^{-2i\omega_0 s_0-2i\omega_1s_1}+H,
\label{a1a2Ansatz}
\end{align*}
where $A$, $B$, $F$, $G$, and $H$ depend solely on $s_1$. Here, we assume that $\zeta$ is at least one order higher than $\epsilon$ so the two time scales make sense: for the oscillations and for the evolution of the amplitudes.

Using the scalings $A,B\propto \sqrt{\epsilon}$, $F,G\propto \epsilon$, and  $H=H_0+\epsilon H_1$, with  $H_0,H_1\propto 1$, Eqs.~\eqref{eqa1} and \eqref{eqa2} are solved order by order in powers of $\epsilon$.
To order $\epsilon^0$, we obtain
\beq
H_0=-\frac{i\beta\zeta\rho_c}{4\pi}\frac{1}{\rho_c+4-i\zeta}.
\eeq
To order $\epsilon^{1/2}$, it results in the linear system 
\beq
\begin{pmatrix}
         -6i(\zeta-2\omega_0) & 32 H_0+3i\beta\zeta \\
	32 H^*_0-3i\beta\zeta & 6i(\zeta+2\omega_0)
\end{pmatrix}
\begin{pmatrix}
         A \\
	B^*
\end{pmatrix}=0,
\eeq
from which it is possible to obtain the relation 
\beq
B^*=\frac{6i(\zeta-2\omega_0)}{32H_0+3i\beta\zeta}A\equiv l A
\eeq
and the frequency
\beq
\omega_0^2=\frac{|32H_0+3i\beta\zeta|^2-36\zeta^2}{4}.
\eeq
We select the positive root; the negative root is equivalent, with the roles of $A$ and $F$ exchanged with those of $B$ and $G$. 
To order $\epsilon$,
\begin{align}
F&=\frac{2\pi}{\rho_{\text{c}}+4-i(\zeta-2\omega_0)}A^2\equiv h_1 A^2,\\
G&=\frac{2\pi}{\rho_{\text{c}}+4-i(\zeta+2\omega_0)}B^2\equiv h_2 B^2,\\
H_1&=\frac{4\pi}{\rho_{\text{c}}+4-i\zeta} \frac{AB}{\epsilon}-
\frac{i\beta\zeta\rho_{\text{c}}/(4\pi)+\rho_{\text{c}}H_0}{\rho_{\text{c}}+4-i\zeta}\\
&\equiv h_3 \frac{AB}{\epsilon}+ h_4.
\label{H1}
\end{align}
Finally, to order $\epsilon^{3/2}$, we obtain the evolution of $A$ in the slow time scale
\beq
\epsilon\frac{dA}{ds_1}=A\left[\epsilon \left(1-i\omega_1-\frac{8}{3}h_4l\right)-\frac{8}{3}\left(h_1+h_3|l|^2\right)|A|^2\right],
\eeq
in which we have already substituted the previous results.

For $\epsilon<0$ ($\rho<\rho_c$), the stationary solution is $A=0$, and the system reaches an isotropic phase where the only contribution to the active stress comes from the imaginary part of $H_0+\epsilon H_1=H_0+\epsilon h_4$. On the contrary, for $\epsilon>0$, a polar phase develops with $A\neq0 $. Then, $F$ and $G$ are nonzero, resulting in a shear stress that oscillates in time. 
For a spherical swimmer ($\beta=0$), $\omega_0= \zeta/2$ and $l=0$.  Using  $\beta=0.7$, which represents \textit{E. coli},  $\omega_0=0.49\zeta+{\textit{O}}(\zeta^2)$.
In this case, $l\approx 0$ and hence $B\approx0$, meaning that the average orientation vector rotates almost harmonically in the same direction as the vorticity. 

To obtain simple analytical expressions, while retaining the main effects, we can neglect $l$. In this case, the stationary solution is $|A|^2=\frac{3\epsilon}{8}\frac{1-i\omega_1}{h_1}$. The latter solution is valid only if it is a real positive number. This condition gives  $\omega_1=-\text{Im}(h_1)/\text{Re}(h_1)$, and therefore the finite value for $|A|^2=\frac{3\epsilon}{8\text{Re}(h_1)}$.  In this approximation $G=0$ and from Eq.~\eqref{aSwimmingSS}, the active shear stress oscillates as
\begin{align}
\Sigma_{xy}^{\text{a}}&=\Sigma_{xy}^{\text{a,mean}}+\Sigma_{xy}^{\text{a,osc}}\sin\left[2(\omega_0+\epsilon \omega_1)s_0-\phi\right],
\end{align}
with amplitude and mean value
\begin{align}
\Sigma_{xy}^{\text{a,osc}}&=3\sigma_0\pi\epsilon/8,\\
\Sigma_{xy}^{\text{a,mean}}&= \frac{\sigma_0\pi\beta\zeta\rho_c}{4\pi(16+\zeta^2+8\rho_c+\rho_c^2)^2}\bigg\{4(16+\zeta^2)(1+\epsilon)\nonumber\\
&\!\!\!\!\!\!\!\!\!\!\!\!\!\!\!\!  +\left[48+32\epsilon+\zeta^2(1+2\epsilon)\right]\rho_c+4(3+\epsilon)\rho_c^2+\rho_c^3 \bigg\}.
 \end{align}

%
%

\begin{figure}
\begin{center}
\includegraphics[width=.8\linewidth]{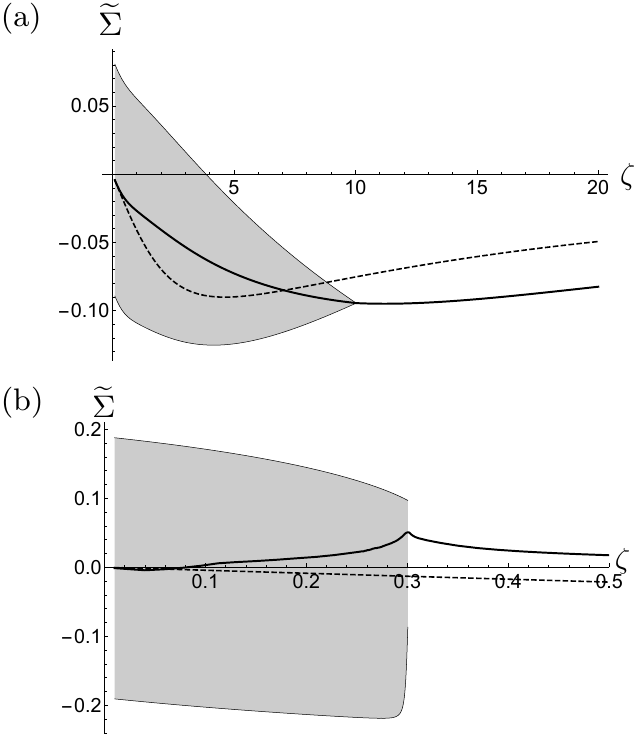}
\end{center}
\caption{
Dimensionless active shear stress $\widetilde\Sigma=\Sigma_{xy}^{\text{a}} g/(D_\text{r} |\sigma_0|)$ as a function of the imposed dimensionless shear rate $\zeta$, for $\beta=0.7$. The solid thick curve shows the mean value, and the shaded region delimitates the possible values when oscillations take place. For comparison, with dashed line, the value without interactions, obtained in Ref.~\cite{saintillan2010dilute}.
(a) Polar aligning interaction with $\eps=0.34$. (b) Axial aligning interaction with $\nu=0.5$. 
}
\label{fig.StressVsGamma}
\end{figure}

\begin{figure}
\begin{center}
\includegraphics[width=0.95\linewidth]{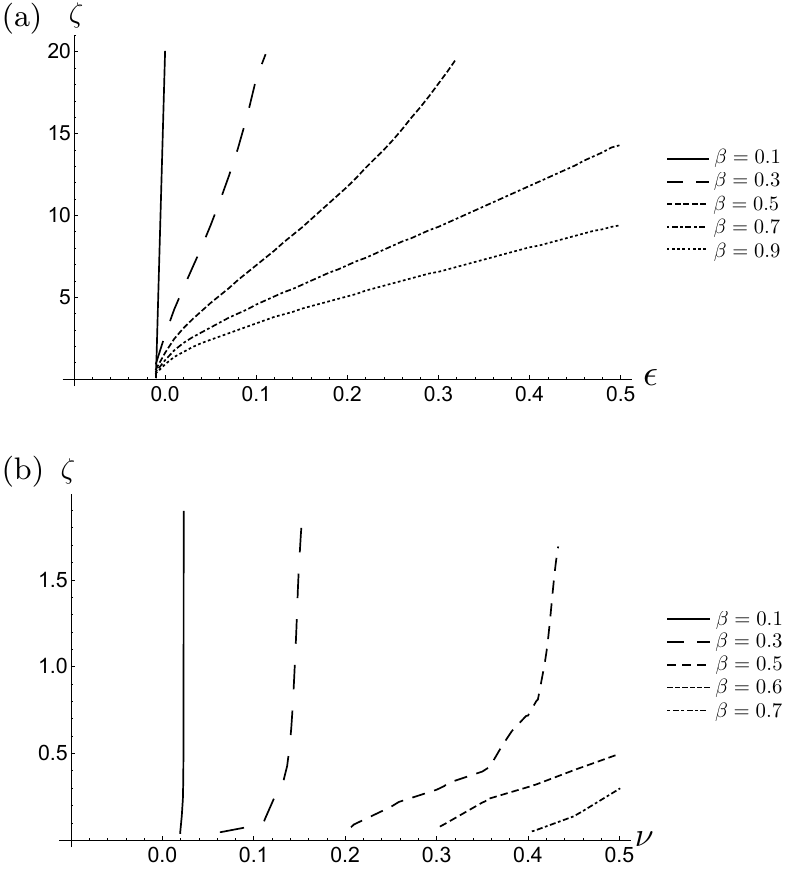}
\end{center}
\caption{(a) Phase diagram in the $\zeta$-$\epsilon$ space for polar aligning interaction and different values of $\beta$. At the right of the respective curves, the suspension orders in a polar phase, while for the rest of the parameter space, the isotropic phase shows up. (b) Same as above in the $\zeta$-$\nu$  space for axial aligning interaction, where the ordered phase is nematic. Notice the change in the scale of the $\zeta$ axis. All curves meet at $\nu=0$ for $\zeta=0$.
}
\label{fig.phasespace}
\end{figure}

Having retained $l$ would give finite values for $B$ and $G$, which manifest in nonharmonic terms in the oscillations of $\Sigma_{xy}^{\text{a}}$.
Numerical solutions of Eqs.~\eqref{FPBn}, truncating at $b_{\pm 5}$, confirm the qualitative behavior obtained in the multiscale scheme, for small values of $\zeta$. However, a different scenario appears for high shear rates, for which the polar phase disappears continuously, as shown in Fig.~\ref{fig.StressVsGamma}. This feature appears only when more than two Fourier modes are considered, otherwise the polar phase remains up to arbitrarily large values of $\zeta$. The maximum shear rate for the polar phase depends on the Bretherton parameter, with a phase diagram presented in Fig.~\ref{fig.phasespace}.

\section{Axial alignment}
For elongated microorganisms such as {\em Bacillus subtilis} or {\em Paramecia}, different observations indicate that they interact axially~\cite{kemkemer2000nematic,kaiser2003coupling,ishikawa2007hydrodynamic,ishikawa2006interaction,zhang2010collective,marchetti2013hydrodynamics}, i.e.\ swimmers with opposite orientations tend to align in an antiparallel way, whereas if they are swimming in a similar direction they align in a parallel manner.
As before, to facilitate the calculations we consider the total aligning case. In this situation,  the alignment is polar if the pre-collisional relative angle is smaller than $\pi/2$ and apolar if it is greater, which is described by the binary collision operator
\begin{multline}
J=g\bigg\{\int_{-\pi/2}^{\pi/2}dw \big[\Psi(\theta+w/2)\Psi(\theta-w/2)\\
+\Psi(\theta+w/2)\Psi(\theta+\pi-w/2)\big]
-\int_{-\pi}^{\pi}\Psi(\theta)\Psi(\theta-w)\bigg\}.
\label{colIntNematic2D}
\end{multline}
The evolution for each of the Fourier amplitudes is given by
\begin{align}
\frac{db_n}{ds}=&-b_n\left(n^2-in\zeta/2+\rho \right)-i\zeta\beta n \left( b_{n-2}+b_{n+2}\right)/4\nonumber\\
&+\pi \sum _{l} b_l b_{n-l}\text{sinc} \left(n\pi/4-l\pi/2\right)\left[1+(-1)^{n-l}\right],
\label{kEqFourierNematicHom}
\end{align}
using the same dimensionless variables as in \eqref{FPBn}. Note that the evolution of the even modes is decoupled from the odd ones, which is a direct consequence of the nematic symmetry present in the collision rule and the swimmer shape.  Equations~\eqref{kEqFourierNematicHom} are solved numerically, truncating at $b_{\pm10}$.

\begin{figure}
\begin{center}
\includegraphics[width=1\linewidth]{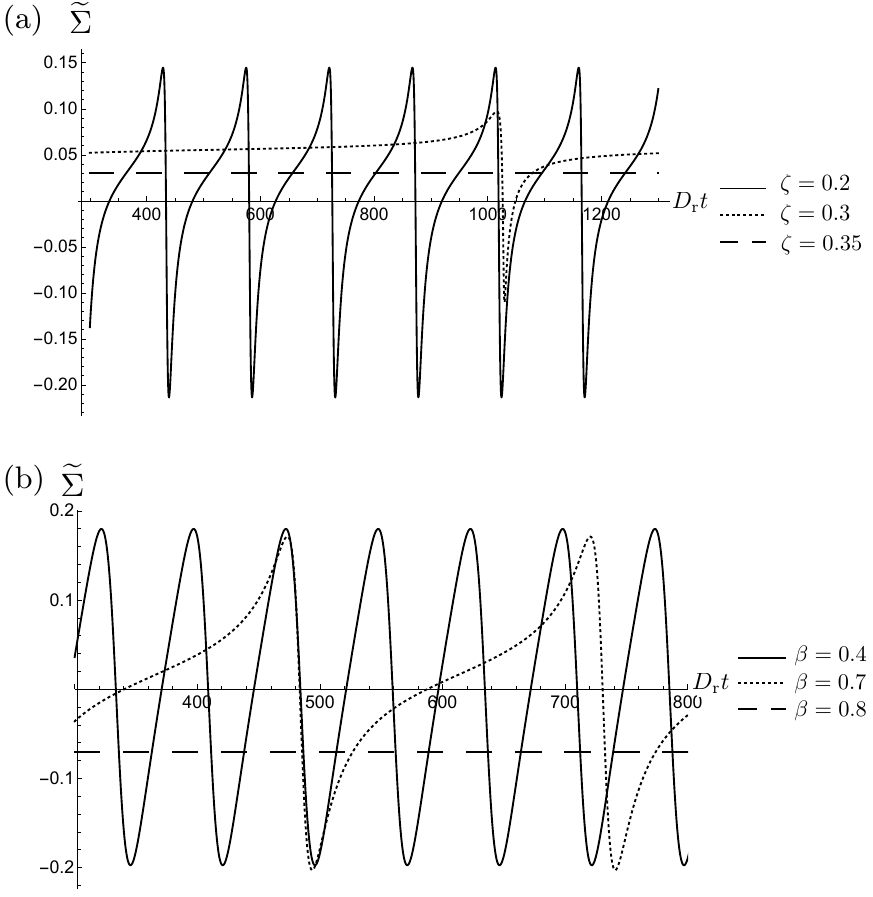}
\end{center}
\caption{Dimensionless active shear stress  $\widetilde\Sigma=\Sigma_{xy}^{\text{a}} g/(D_\text{r} |\sigma_0|)$ versus time for axial aligning interaction  with $\nu=0.5$, when approaching the transition at high shear rates. 
(a) $\beta=0.7$, and different shear rates. (b) $\zeta=0.1$, and different Bretherton coefficients. Notice that in the two panels, the horizontal axes do not span the same amount of time.}
\label{fig.comparisonShape}
\end{figure}

We define the bifurcation parameter $\nu=\rho (4/\pi-1)-4$,
such that for $\nu=0$, there is a transition from the isotropic to nematic phase when $\zeta=0$. This time the critical density is four times bigger than in the polar case, $\rho_c^{\text{nematic}}=4\rho_c$.
When $\zeta\neq0$, the nematic phase leads to a similar qualitative behavior to the polar case, namely, an oscillating active shear stress. However, important differences appear. First, the average orientation vector is always zero, indicating that no polar order appears.  
Second, the transition from the ordered to the isotropic phase takes place for very small  shear rates. It is necessary to consider large values of $\nu$ (increase the density) to remain in the nematic phase for finite values of $\zeta$, particularly for very slender swimmers (see Fig.~\ref{fig.phasespace}, bottom). 
As a consequence that $\nu$ is not necessarily small, there is no full enslaving of the higher modes to the critical one, $b_{\pm 2}$, and the oscillations become non-harmonic, contrary to the polar aligning case, where the oscillations remain roughly harmonic for all parameters. 
Notably, due to the anharmonicities, the average active shear stress can be positive instead of negative, meaning that pushers interacting axially increase the effective viscosity (see Fig.~\ref{fig.StressVsGamma}, bottom).
Finally, at the transition from the nematic to the isotropic phase,  the amplitude of the oscillation vanishes abruptly, and the period diverges with extremely nonharmonic dynamics (see Figs.~\ref{fig.StressVsGamma} and \ref{fig.comparisonShape}). These are characteristic features of the limit cycle disappearing via a homoclinic bifurcation~\cite{vitt1996theory, strogatz2018nonlinear}.

\section{Discussion}
In this article we have reviewed the nonideal rheology of bacterial suspensions in the semi-dilute regime. Considering both polar and axial aligning collisions, we have obtained that the short-range interactions lead to an oscillatory active shear stress, in response to a constant shear rate. 
In both cases we found that there is a competition between the rotation imposed by the shear rate and the short-range interactions. When the shear rate is large, 
the ordered phase is lost in a continuous  or discontinuous transition, for polar or axial interactions, respectively.

To our knowledge, just one experiment has measured with high precision the shear stress in bacterial suspensions~\cite{lopez2015turning}. There, using an \textit{E. coli} suspension, they observed a large reduction of viscosity (up to a superfluid state) for volume fractions in the range $0.11\%\leq\phi\leq0.67\%$, which are higher than the critical density predicted in this model for polar aligning interactions ($\phi_{\text{c}}\approx0.1\%$). These volume fractions correspond to the bifurcation parameter in the range $-0.03\leq\epsilon\leq4.9$ or $-3.03\leq\nu\leq1.9$. The experiments used a  high-precision Taylor-Couette rheometer subjected to a small shear rate  $\dgam=\SI{0.04}{\second^{-1}}$ ($\zeta\approx 3.2$). Although these experiments fell well inside the polar phase (see Fig.~\ref{fig.phasespace}), no oscillation in the shear stress was observed. This can be understood by at least three facts. First, the period of the oscillations is $T\sim 2\pi/\dgam\approx\SI{150}{\second}$, which is larger than the $\SI{30}{\second}$ that the experiment lasted and hence there was no time to observe the oscillation. Second, in an extended system, it takes a long time for the different domains to synchronize. Then, the rheometer measures an incoherent sum of oscillatory stresses. Finally, the buildup time of the polar phase scales as $1/(D_\text{r}\epsilon)$, ranging between $\SI{8}{\second}$ and $\SI{100}{\second}$, which limits also the possibility to observe the oscillations in the earlier experiment. 

In this article, we have focused on pusher swimmers, for which was predicted a reduction of the viscosity in the dilute limit~\cite{saintillan2010dilute}. However, two remarkable effects appear. With axial alignment, the average viscosity increases instead. And the shear stress oscillations involve also an increase of the viscosity during half of the oscillation period. Finally, this implies that pullers should reduce the suspension viscosity under appropriate conditions.

The collision rules employed in this article are totally aligning, and we did not consider any threshold for the relative angle to select which bacteria  interact, for example. Nevertheless, we expect the presented results to remain qualitatively the same by relaxing these conditions. 
For axial aligning collisions, the oscillation period varies quite dramatically near the homoclinic transition, for example, varying the Bretherton coefficient, as shown in Fig.~\ref{fig.comparisonShape}, bottom. Then, in a  suspension with a diversity of shapes, interesting segregation or synchronization effects can appear. Moreover, a more realistic situation could be accounted for by mixing the two collision rules, with  collision factors $g^{\text{polar}}$ and $g^{\text{axial}}$. In this regard, some intuition can be obtained from the study of active polar liquid crystals of slender rods ($\beta \rightarrow 1$)~\cite{forest2014rheological}, where these two interactions are dealt with via intermolecular potentials~\cite{masao1981molecular}, although there is not a clear relation between the nematic and polar strengths of the potential and the bacterial density.
Further  studies are needed to understand these effects.

For the three-dimensional case, the distribution function depends on the two spherical angles that parametrize the directors. An expansion of the distribution function in spherical harmonics $Y_{lm}$ would allow one to perform an analogous analysis as in this article. Again, the average orientation vector depends on the $l=1$ modes and the active stress tensor on the $l=2$ modes. Symmetry arguments dictate that the same nonlinear mode couplings will take place, for example, coupling only even modes for the axial alignment case. Hence, although the calculations are much more involved, it is expected that similar rotating polar and nematic phases develop, which will imply that the active stresses will oscillate in time. The eventual end of these phases when increasing the shear rate, in the form of supercritical or homoclinic bifurcations, depends on the explicit expressions for the mode equations and is a subject of future research.

\begin{acknowledgments}
We thank M. Clerc for his comments on the theory of dynamical systems and E. Clement for enlightening discussions on the bacterial rheology. This research was supported by the Fondecyt Grant No.~1180791 and the Millennium Nucleus ``Physics of active matter'' of the Millennium Scientific Initiative of the Ministry of Economy, Development and Tourism (Chile). M.G.~acknowledges the Conicyt PFCHA Magister Nacional Scholarship 2016-22162176. 
\end{acknowledgments}

\end{document}